\documentclass[10pt]{article}

\usepackage{graphicx}
\usepackage{psfrag}
\usepackage{subfigure}
\usepackage{amsmath,amssymb}
\usepackage{fullpage}

\def \bv{{\bf v}}
\def \bc{{\bf c}}
\def \bm{{\bf m}}

\newcommand{\comment}[1]{}

\title{Improved Examples of \\Non-Termination for Ruppert's Algorithm}
\author{Alexander Rand\thanks{Institute for Computational Engineering and Sciences, University of Texas at Austin, \texttt{arand@ices.utexas.edu}}}
\date{\today}

\begin{document}

\maketitle

\begin{abstract}
Improving the best known examples, two planar straight-line graphs which cause the non-termination of Ruppert's algorithm for a minimum angle threshold as low as $\alpha \gtrapprox 29.06^\circ$ are given. 
\end{abstract}

\section*{Introduction}

Given a planar straight-line graph (PSLG), Ruppert's algorithm \cite{Ru95} produces a conforming Delaunay triangulation satisfying a minimum angle bound.  The standard analysis \cite{Ru95,Sh02} demonstrates that when the input contains no angles smaller than $60^\circ$ Ruppert's algorithm produces a size-optimal mesh (up to a constant factor) for any minimum angle bound $\alpha \lessapprox 20.7^\circ$.  A more detailed analysis can slightly improve this restriction to $\alpha \lessapprox 22.2^\circ$ for non-acute input~\cite{Ra10} and an additional (very mild) assumption further improves the guarantee to $26.5^\circ$~\cite{MPW03}.  

In practice, Ruppert's algorithm succeeds for substantially larger minimum angle bounds than those guaranteed by the theory.  Ruppert observed that the minimum angle reaches $30^\circ$ during typical runs of the algorithm~\cite{Ru95}.  In a number of experiments Shewchuk found the algorithm to terminate for $\alpha \lessapprox 33.8^\circ$~\cite{Sh96}.  Certain modifications of the vertex insertion procedure further improve the algorithm suggesting that this constraint can be improved to possibly $40^\circ$ or more~\cite{EU09}.

Pav demonstrated an example of non-termination of Ruppert's algorithm on a simple non-acute PSLG for any $\alpha >30^\circ$~\cite{Pa03}.  This example combined with the fact that the analysis of Ruppert's algorithm \emph{on point sets} breaks down at $30^\circ$ led to a natural conjecture that Ruppert's algorithm terminates and produces a well-graded mesh for all $\alpha < 30^\circ$. While some examples corroborated this idea~\cite{Ra10}, a recent example shows non-termination for $\alpha$ as low as $\approx 29.51^\circ$~\cite{Ra11}. 

We give two examples that improve upon those given in \cite{Ra11}.  The first example produces non-termination of Ruppert's algorithm for $\alpha \gtrapprox 29.10^\circ$ and has a minimum input angle of about $87^\circ$.  The second example gives a slight improvement, $\alpha \gtrapprox 29.06^\circ$ but requires a $60^\circ$ input angle. 

First, some notation is defined.  The line-segment between endpoints $\bv_1$ and $\bv_2$ is denoted $\overline{\bv_1\bv_2}$ and the triangle with vertices $\bv_1$, $\bv_2$, and $\bv_3$ is $\triangle \bv_1\bv_2\bv_3$.  Let $\angle \bv_1\bv_0\bv_2$ be the angle at vertex $\bv_0$ between line-segments $\overline{\bv_0\bv_1}$ and $\overline{\bv_0\bv_2}$.    

\section*{Example 1}

\begin{figure}
\centering
\includegraphics[height=.4\textwidth]{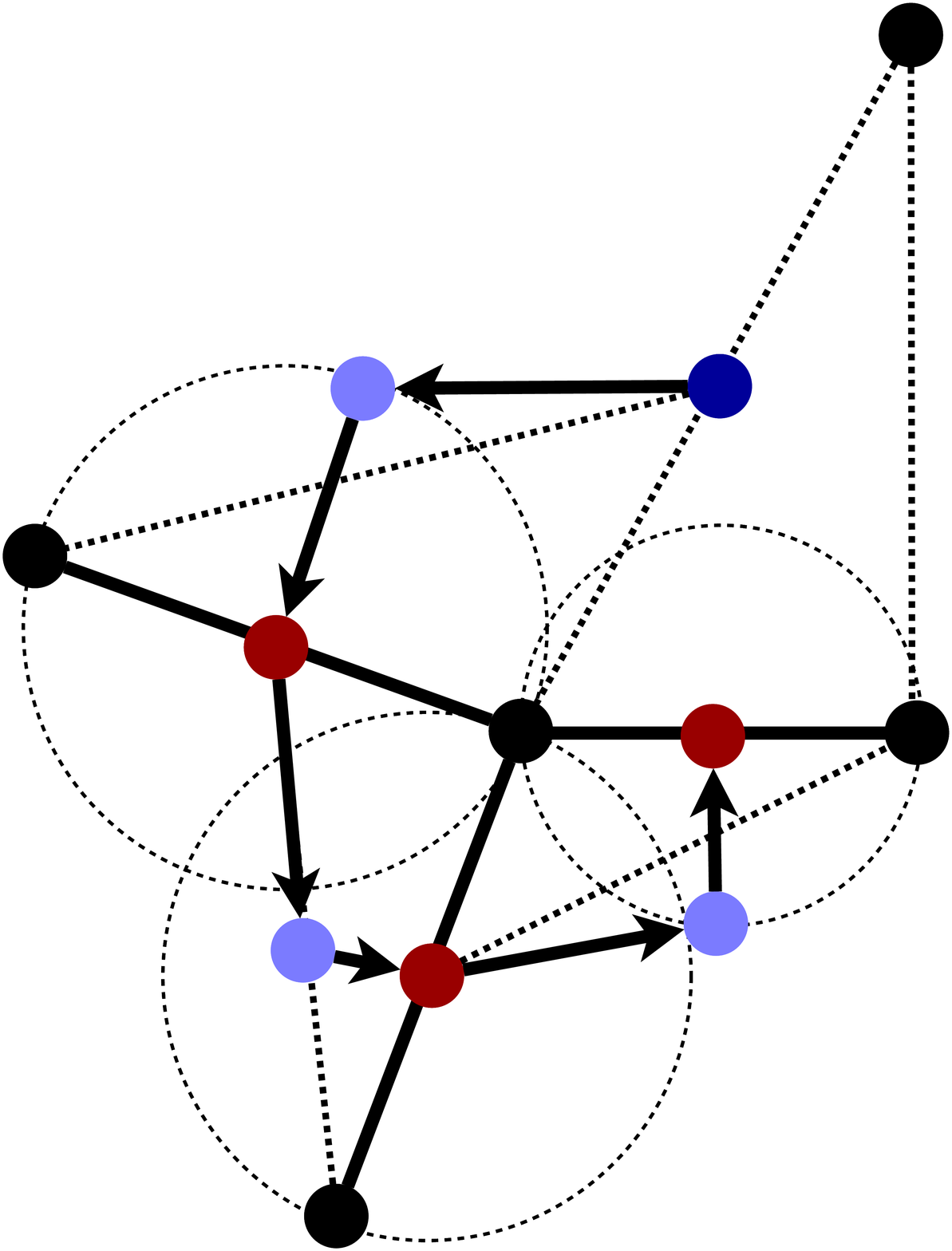}
\hspace{.5in}
\psfrag{v0}{$\bv_0$}
\psfrag{v1}{$\bv_1$}
\psfrag{v2}{$\bv_2$}
\psfrag{v3}{$\bv_3$}
\psfrag{v4}{$\bv_4$}
\psfrag{c1}{$\bc_1$}
\psfrag{c2}{$\bc_2$}
\psfrag{c3}{$\bc_3$}
\psfrag{c4}{$\bc_4$}
\psfrag{m1}{$\bm_1$}
\psfrag{m2}{$\bm_2$}
\psfrag{m3}{$\bm_3$}
\includegraphics[height=.4\textwidth]{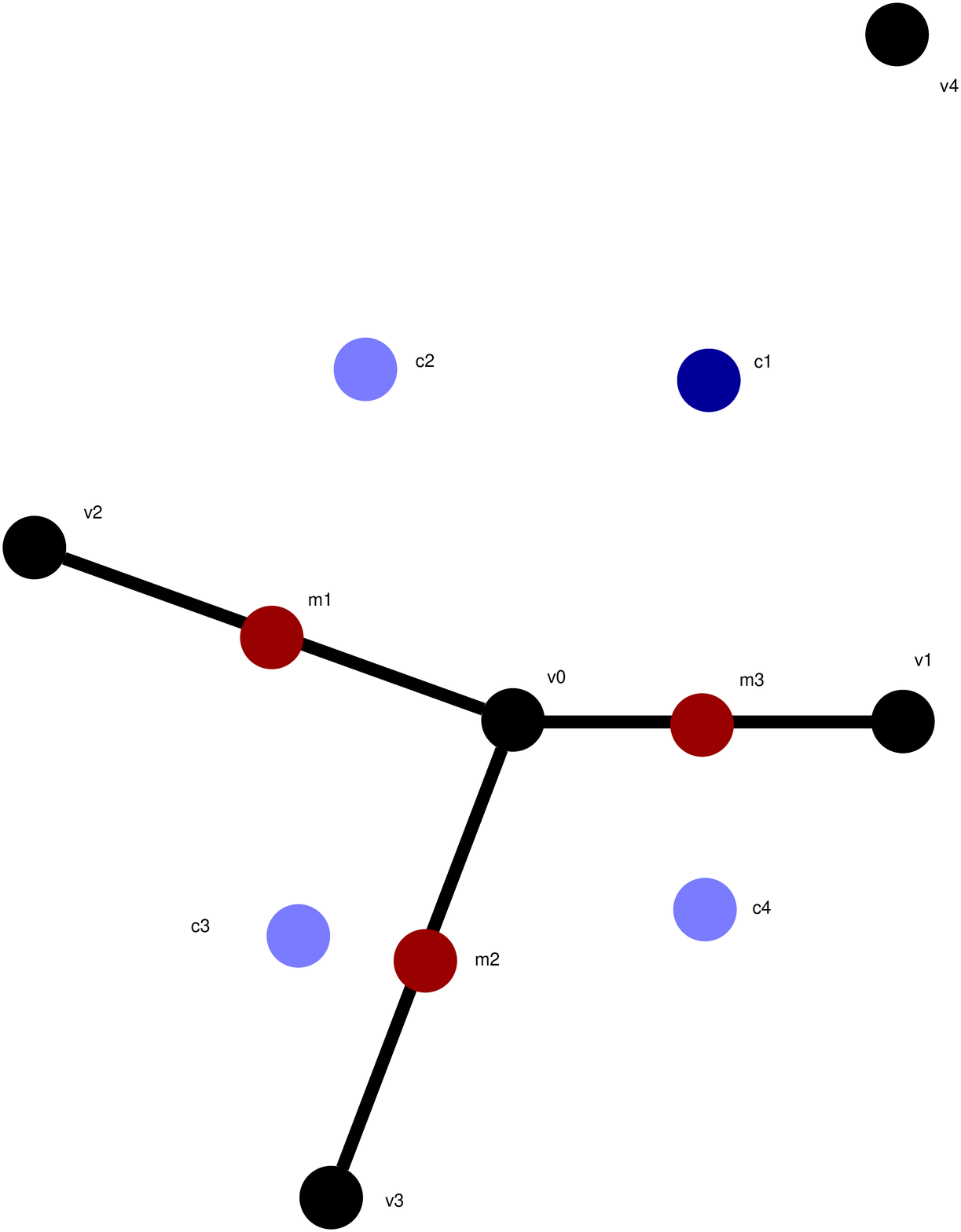}
\caption{Example 1.  The black vertices and segments comprise the input.  Blue vertices are circumcenters of skinny triangles: the dark blue vertex is inserted into the triangulation while the light blue vertices are rejected for encroaching segments.  The red vertices are midpoints of encroached segments which are inserted by Ruppert's algorithm. (left) The sequence of vertices considered for insertion by Ruppert's algorithm with the relevant skinny triangles and encroached diametral circles.  (right) Labels for each of the vertices.}\label{fg:counterExNew}
\end{figure}

This example PSLG consists of five input vertices and three adjacent input segments.  By carefully constructing the input (as described below) Ruppert's algorithm inserts a circumcenter, followed by three consecutive circumcenters that yield to midpoints of the segments.  The result is a configuration that is similar to the input but exactly half the size and thus Ruppert's algorithm can repeat this cycle indefinitely.  Depicted in Figure~\ref{fg:counterExNew} the sequence involves four vertex insertions: (1) the circumcenter $\bc_1$ of skinny triangle $\triangle \bv_0\bv_1\bv_4$ is inserted, (2) the circumcenter $\bc_2$ of skinny triangle $\triangle \bv_0\bv_2\bc_1$ encroaches segment $\overline{\bv_0\bv_2}$ which causes the insertion of midpoint $\bm_1$, (3) the circumcenter $\bc_3$ of skinny triangle $\triangle \bv_0\bv_3\bm_1$ encroaches segment $\overline{\bv_0\bv_3}$ which causes the insertion of midpoint $\bm_2$, and finally (4) the circumcenter of skinny triangle $\triangle \bv_0\bv_1\bm_2$ encroaches segment $\overline{\bv_0\bv_1}$ which causes the insertion of midpoint $\bm_3$.  

\paragraph{Construction}

The following construction describes how to place the input vertices ($\bv_0, \ldots, \bv_4$) to achieve non-termination for the smallest possible angle threshold.  We begin by fixing $\gamma_1$ which serves as the smallest angle for a number of skinny triangles in the example.  Then the placement of the vertices is determined by the following steps; recall Figure~\ref{fg:counterExNew}.
\begin{enumerate}
\item Begin with a segment between two vertices $\bv_0$ and $\bv_1$.  
\item Let $\bm_3$ be the midpoint of $\overline{\bv_0\bv_1}$.  
\item Place $\bv_4$ such that $\angle \bv_4\bv_1\bv_0 = 90^\circ$ and $\angle \bv_1\bv_4\bv_0 = \gamma_1$. 
\item Let $\bc_1$ denote the circumcenter of $\triangle \bv_0\bv_1\bv_4$.   
\item Select $\bv_2$ such that the circumcenter of $\triangle \bv_0\bv_2\bc_1$, denoted $\bc_2$, lies on the diametral ball of $\overline{\bv_0\bv_2}$.  
\item Let $\bm_1$ denote the midpoint of $\overline{\bv_0\bv_2}$.  
\item Select $\bm_2$ so that the circumcenter of $\triangle \bv_0\bv_1\bm_2$, denoted $\bc_4$, lies on the diametral ball of $\overline{\bv_0\bv_1}$ and the opposite side of $\overline{\bv_0\bv_1}$ as $\bc_1$.  
\item Define $\bv_3$ so that $\bm_2$ is the midpoint of $\overline{\bv_0\bv_3}$. 
\item Let $\bc_3$ denote the circumcenter of $\triangle \bv_0\bv_3\bm_1$.  
\end{enumerate}

When $\alpha > \gamma_1$ the construction ensures that triangles $\triangle \bv_0\bv_1\bv_4$ and (following the insertion of $\bc_1$) $\triangle \bv_0\bv_2\bc_1$ are split by Ruppert's algorithm.  Moreover, if $\bm_2$ is inserted, then $\triangle \bv_0\bv_1\bm_2$ is skinny.  Also, by construction, if/when $\bc_2$ and $\bc_4$ are inserted, they encroach $\overline{\bv_0\bv_2}$ and $\overline{\bv_0\bv_1}$, respectively, resulting in the insertion of $\bm_1$ and $\bm_3$.  

For non-termination to occur, $\triangle \bv_0\bv_3\bm_1$ must also be skinny and then its circumcenter $\bc_3$ must encroach $\overline{\bv_0\bv_3}$.  When $\gamma_1 \in [25,30]$, $\bc_3$ always encroaches $\overline{\bv_0\bv_3}$ and this encroachment is not `sharp' unlike the encroachment of $\overline{\bv_0\bv_1}$ and $\overline{\bv_0\bv_2}$ which is constructed to occur on the boundary of the diametral ball.  

Letting $\gamma_2:= \angle \bv_0\bv_3\bm_1$, non-termination occurs for $\alpha > \max(\gamma_1,\gamma_2)$. Thus, the best example will minimize $\max(\gamma_1,\gamma_2)$.  Since $\gamma_2$ is a function of $\gamma_1$ (i.e., given $\gamma_1$, performing the construction yields $\gamma_2$), this optimization need only be performed over a one-parameter family.  Numerically we find that $\gamma_1 = \gamma_2 \approx 29.10^\circ$ produces the smallest required threshold.

\paragraph{Remarks}

\begin{itemize}
\item The PSLG input constructed is slightly acute: $\angle \bv_2\bv_0\bv_3 \approx 87.3^\circ$.  Restricting the construction to ensure that all input angles are larger than $90^\circ$ yields an example of non-termination for $\alpha > 30^\circ$, i.e., no better than Pav's original example.
\item There appears to be some `slack' in the construction since $\bc_3$ lies well inside the diametral ball of $\overline{\bv_0\bv_3}$.  However, there is no perturbation of $\bv_2$ or $\bv_3$ that yields a valid encroachment sequence and improves on the needed angle threshold.  
\item Removing symmetry is an important part of the construction.  The original example~\cite{Ra10} contained four similar configurations between adjacent segments and the first improvement~\cite{Ra11} broke the symmetry leaving two similar constructions.  This second improved example goes one step further by eliminating all symmetry.
\item The requirement $\angle \bv_0\bv_1\bv_4 = 90^\circ$ is essential to ensure that $\bc_1$ lies in the correct location for subsequent iterations.  
\end{itemize}

\section*{Example 2}

\begin{figure}
\centering
\includegraphics[height=.4\textwidth]{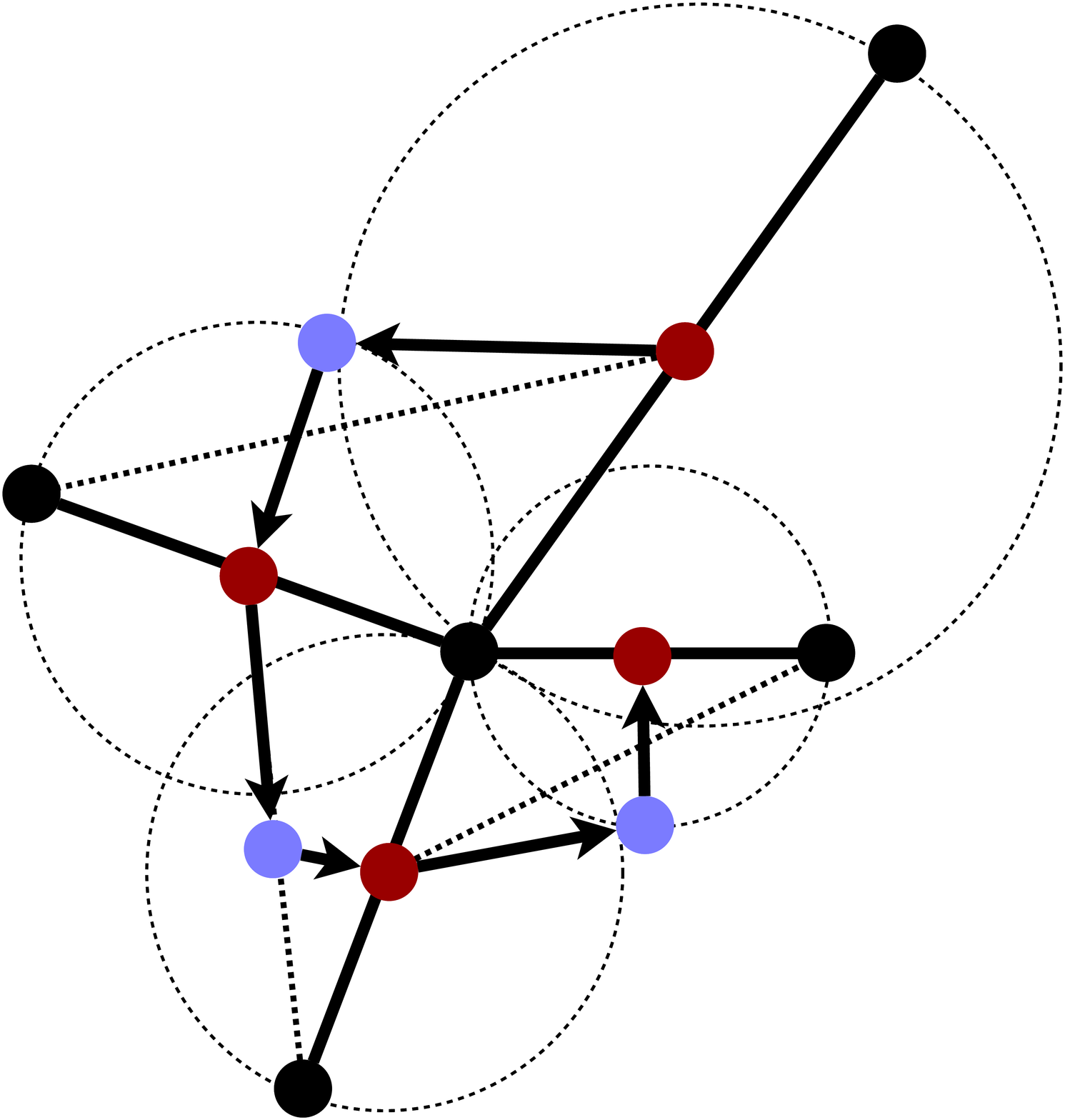}
\hspace{.5in}
\psfrag{v0}{$\bv_0$}
\psfrag{v1}{$\bv_1$}
\psfrag{v2}{$\bv_2$}
\psfrag{v3}{$\bv_3$}
\psfrag{v4}{$\bv_4$}
\psfrag{c1}{$\bc_1$}
\psfrag{c2}{$\bc_2$}
\psfrag{c3}{$\bc_3$}
\psfrag{c4}{$\bc_4$}
\psfrag{m0}{$\bm_0$}
\psfrag{m1}{$\bm_1$}
\psfrag{m2}{$\bm_2$}
\psfrag{m3}{$\bm_3$}
\includegraphics[height=.4\textwidth]{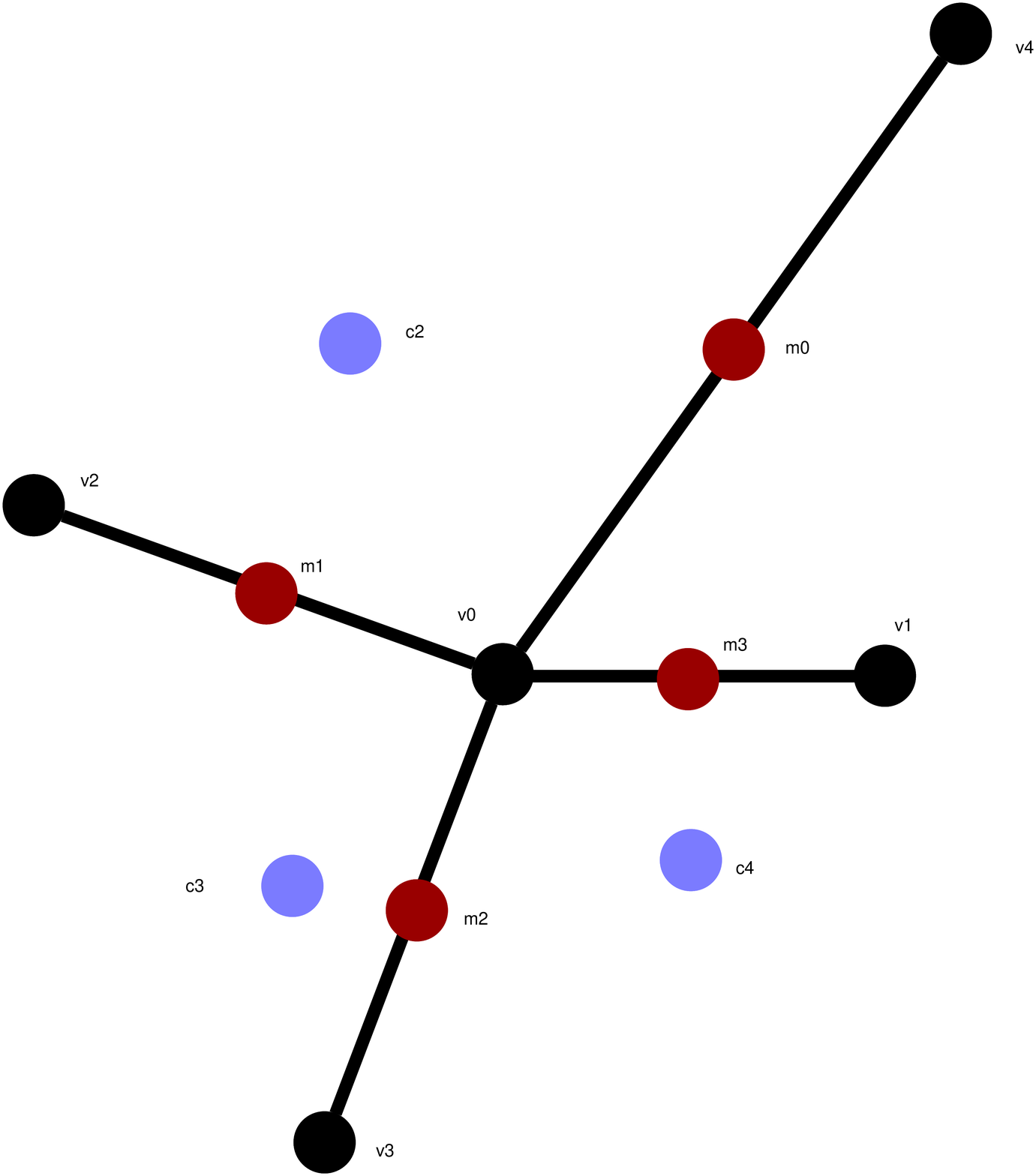}
\caption{Example 2. The black vertices and segments comprise the input.  Light blue vertices are circumcenters of skinny triangles which are rejected for encroaching segments.  The red vertices are midpoints of encroached segments which are inserted by the algorithm. (left) The sequence of vertices considered for insertion by Ruppert's algorithm with the relevant skinny triangles and encroached diametral circles. (right) Labels for each of the vertices.}\label{fg:counterExSixty}
\end{figure}

As noted in the previous remark, there is little flexibility in the position of the unconnected vertex $\bv_4$.  If segment $\overline{\bv_0\bv_4}$ is included in the input, some flexibility is gained since the midpoint (denoted $\bm_0$) creates a similar configuration for the next cycle of the algorithm even if $\angle \bv_0\bv_1\bv_4 \neq 90^\circ$.  Essentially the same sequence of vertex insertions is caused by Ruppert's algorithm except now $\bv_1$ encroaches segment $\overline{\bv_0\bv_4}$; see Figure~\ref{fg:counterExSixty}.  Then, $\bc_2$, $\bm_1$, etc. are inserted in a similar fashion where $\bm_0$ replaces $\bc_1$ in the sequence. The new construction and requirements on $\alpha$ are now described.

\paragraph{Construction}

The construction is nearly identical to Example 1.  In addition to adding the segment $\overline{\bv_0\bv_4}$ to the input, the placement of $\bv_4$ is slightly different.  Steps 3 and 4 of the construction in Example 1 are replaced with the following steps.  
\begin{enumerate}
\item[3*.] Place $\bv_4$ such that $\angle \bv_4\bv_0\bv_1 = 60^\circ$ and $\angle \bv_1\bv_4\bv_0 = \gamma_1$. 
\item[4*.] Let $\bm_0$ denote the midpoint of segment $\overline{\bv_0\bv_4}$.   
\end{enumerate}
The later steps of the construction in Example 1 are all identical with a single exception: midpoint $\bm_0$ replaces circumcenter $\bc_1$ in all cases.  

As in the first example, $\gamma_1$ can be selected to match the smallest angle $\gamma_2$ of the final skinny triangle (recall $\gamma_2 = \angle \bv_0\bv_3\bm_1$). The result is an input for which Ruppert's algorithm does not terminate for all $\alpha \gtrapprox 29.06^\circ$.  

\paragraph{Remarks}
\begin{itemize}
\item Example 2 more clearly identifies the challenge in utilizing the `slack' in positioning $\bc_3$ to improve the example.  Such modification seeks to increase $\angle \bv_2\bv_0\bv_3$ at the expense of $\angle \bv_1\bv_0\bv_4$ which would violate our restriction that input angles are larger than $60^\circ$.  
\item Relaxing the $60^\circ$ restriction on input angles to $51^\circ$ (which corresponds to the best known counterexample~\cite{Pa03}) would improve our result to $\alpha \gtrapprox 28.46^\circ$.  Allowing a single $45^\circ$ input angle (the value at which trivial alternating midpoint insertion can occur) pushes the bound to $\alpha \gtrapprox 28.00^\circ$
\end{itemize}

\section*{Conclusion}

Examples of the non-termination of Ruppert's algorithm serve an important role in the development of a sharp analysis of the algorithm.  The PSLG inputs we have constructed improve the best known examples of non-termination due to the size of the minimum angle threshold selected.  

If a refined analysis of Ruppert's algorithm is going to yield guarantees that more closely resemble its behavior in practice, these counterexamples demonstrate why additional mild assumptions on the input or the algorithm must be considered.  Non-acute input (rather than admitting $60^\circ$ input angles), groomed input so that adjacent segments have equal length, restrictions on queue ordering, and non-circumcenter Steiner vertices are all possible candidates which have been used to improve Delaunay refinement in theory and/or practice; e.g., \cite{MPW03,Un04,Ra08,HMP06,Ra10}.  Without explicitly utilizing any of these modifications, any extension of the analysis must be limited by the $\alpha \approx 29.06^\circ$ example.

\bibliographystyle{abbrv}
\bibliography{ruppert}

\begin{thebibliography}{10}

\bibitem{EU09}
H.~Erten and A.~\"Ung\"or.
\newblock Quality triangulations with locally optimal {S}teiner points.
\newblock {\em SIAM J. Sci. Comput.}, 31:2103--2130, 2009.

\bibitem{HMP06}
B.~Hudson, G.~Miller, and T.~Phillips.
\newblock {S}parse {V}oronoi refinement.
\newblock In {\em Proc. 15th Int. Meshing Roundtable}, pages 339--356, 2006.

\bibitem{MPW03}
G.~Miller, S.~Pav, and N.~Walkington.
\newblock When and why {R}uppert's algorithm works.
\newblock In {\em Proc. 12th Int. Meshing Roundtable}, pages 91--102, 2003.

\bibitem{Pa03}
S.~Pav.
\newblock {\em Delaunay Refinement Algorithms}.
\newblock PhD thesis, Carnegie Mellon, 2003.

\bibitem{Ra08}
A.~Rand.
\newblock Reordering {R}uppert's algorithm.
\newblock In {\em Proc. 18th Fall Workshop Comput. Geom.}, pages 43--44, 2008.

\bibitem{Ra10}
A.~Rand.
\newblock On the termination of {R}uppert's algorithm.
\newblock In {\em Res. Notes 19th Int. Meshing Roundtable}, 2010.

\bibitem{Ra11}
A.~Rand.
\newblock On the non-termination of {R}uppert's algorithm.
\newblock {\tt arXiv:1101.1071v1 [cs.CG]}, 2011.

\bibitem{Ru95}
J.~Ruppert.
\newblock A {D}elaunay refinement algorithm for quality 2-dimensional mesh
  generation.
\newblock {\em J. Algorithms}, 18(3):548--585, 1995.

\bibitem{Sh96}
J.~Shewchuk.
\newblock Triangle: {E}ngineering a {2D} quality mesh generator and {D}elaunay
  triangulator.
\newblock In M.~Lin and D.~Manocha, editors, {\em Applied Computational
  Geometry Towards Geometric Engineering}, pages 203--222. Springer, 1996.

\bibitem{Sh02}
J.~Shewchuk.
\newblock Delaunay refinement algorithms for triangular mesh generation.
\newblock {\em Comput. Geom.}, 22(1--3):86--95, 2002.

\bibitem{Un04}
A.~\"Ung\"or.
\newblock Off-centers: A new type of {S}teiner points for computing
  size-optimal quality-guaranteed {D}elaunay triangulations.
\newblock In {\em Proc. 6th Latin Amer. Symp. Theor. Inform.}, pages 152--161,
  2004.

\end{thebibliography}

\end{document}